\documentclass[prd,twocolumn,preprintnumbers,floatfix,letterpaper,showpacs,showkeys,superscriptaddress,nofootinbib]{revtex4}

\usepackage[dvipdf]{graphicx}
\usepackage{bm}

\begin{document}

\title{
Measurement of neutrino velocity with the MINOS detectors\\
and NuMI neutrino beam
}

\newcommand{\corresp}{Corresponding Author.}
\newcommand{\Cambridge}{Cavendish Laboratory, University of Cambridge, Madingley Road, Cambridge CB3 0HE, United Kingdom}
\newcommand{\FNAL}{Fermi National Accelerator Laboratory, Batavia, Illinois 60510, USA}
\newcommand{\RAL}{Rutherford Appleton Laboratory, Chilton, Didcot, Oxfordshire, OX11 0QX, United Kingdom}
\newcommand{\UCL}{Department of Physics and Astronomy, University College London, Gower Street, London WC1E 6BT, United Kingdom}
\newcommand{\Caltech}{Lauritsen Laboratory, California Institute of Technology, Pasadena, California 91125, USA}
\newcommand{\ANL}{Argonne National Laboratory, Argonne, Illinois 60439, USA}
\newcommand{\Athens}{Department of Physics, University of Athens, GR-15771 Athens, Greece}
\newcommand{\NTUAthens}{Department of Physics, National Tech. University of Athens, GR-15780 Athens, Greece}
\newcommand{\Benedictine}{Physics Department, Benedictine University, Lisle, Illinois 60532, USA}
\newcommand{\BNL}{Brookhaven National Laboratory, Upton, New York 11973, USA}
\newcommand{\CdF}{APC -- Universit\'{e} Paris 7 Denis Diderot, 10, rue Alice Domon et L\'{e}onie Duquet, F-75205 Paris Cedex 13, France}
\newcommand{\Cleveland}{Cleveland Clinic, Cleveland, Ohio 44195, USA}
\newcommand{\GEHealth}{GE Healthcare, Florence South Carolina 29501, USA}
\newcommand{\Harvard}{Department of Physics, Harvard University, Cambridge, Massachusetts 02138, USA}
\newcommand{\HolyCross}{Holy Cross College, Notre Dame, Indiana 46556, USA}
\newcommand{\IIT}{Physics Division, Illinois Institute of Technology, Chicago, Illinois 60616, USA}
\newcommand{\Indiana}{Indiana University, Bloomington, Indiana 47405, USA}
\newcommand{\ITEP}{High Energy Experimental Physics Department, Institute of Theoretical and Experimental Physics, 
  B. Cheremushkinskaya, 25, 117218 Moscow, Russia}
\newcommand{\JMU}{Physics Department, James Madison University, Harrisonburg, Virginia 22807, USA}
\newcommand{\LASL}{Nuclear Nonproliferation Division, Threat Reduction Directorate, Los Alamos National Laboratory, Los Alamos, New Mexico 87545, USA}
\newcommand{\Lebedev}{Nuclear Physics Department, Lebedev Physical Institute, Leninsky Prospect 53, 117924 Moscow, Russia}
\newcommand{\LLL}{Lawrence Livermore National Laboratory, Livermore, California 94550, USA}
\newcommand{\MIT}{Lincoln Laboratory, Massachusetts Institute of Technology, Lexington, Massachusetts 02420, USA}
\newcommand{\Minnesota}{University of Minnesota, Minneapolis, Minnesota 55455, USA}
\newcommand{\Crookston}{Math, Science and Technology Department, University of Minnesota -- Crookston, Crookston, Minnesota 56716, USA}
\newcommand{\Duluth}{Department of Physics, University of Minnesota -- Duluth, Duluth, Minnesota 55812, USA}
\newcommand{\Oxford}{Subdepartment of Particle Physics, University of Oxford,  Denys Wilkinson Building, Keble Road, Oxford OX1 3RH, United Kingdom}
\newcommand{\Pittsburgh}{Department of Physics and Astronomy, University of Pittsburgh, Pittsburgh, Pennsylvania 15260, USA}
\newcommand{\IHEP}{Institute for High Energy Physics, Protvino, Moscow Region RU-140284, Russia}
\newcommand{\RoyalH}{Physics Department, Royal Holloway, University of London, Egham, Surrey, TW20 0EX, United Kingdom}
\newcommand{\Carolina}{Department of Physics and Astronomy, University of South Carolina, Columbia, South Carolina 29208, USA}
\newcommand{\SLAC}{Stanford Linear Accelerator Center, Stanford, California 94309, USA}
\newcommand{\Stanford}{Department of Physics, Stanford University, Stanford, California 94305, USA}
\newcommand{\Sussex}{Department of Physics and Astronomy, University of Sussex, Falmer, Brighton BN1 9QH, United Kingdom}
\newcommand{\TexasAM}{Physics Department, Texas A\&M University, College Station, Texas 77843, USA}
\newcommand{\Texas}{Department of Physics, University of Texas, 1 University Station, Austin, Texas 78712, USA}
\newcommand{\TechX}{Tech-X Corporation, Boulder, Colorado 80303, USA}
\newcommand{\Tufts}{Physics Department, Tufts University, Medford, Massachusetts 02155, USA}
\newcommand{\UNICAMP}{Universidade Estadual de Campinas, IF-UNICAMP, CP 6165, 13083-970, Campinas, SP, Brazil}
\newcommand{\USP}{Instituto de F\'{i}sica, Universidade de S\~{a}o Paulo,  CP 66318, 05315-970, S\~{a}o Paulo, SP, Brazil}
\newcommand{\Washington}{Physics Department, Western Washington University, Bellingham, Washington 98225, USA}
\newcommand{\WandM}{Department of Physics, College of William \& Mary, Williamsburg, Virginia 23187, USA}
\newcommand{\Wisconsin}{Physics Department, University of Wisconsin, Madison, Wisconsin 53706, USA}
\newcommand{\deceased}{Deceased.}

\affiliation{\ANL}
\affiliation{\Athens}
\affiliation{\Benedictine}
\affiliation{\BNL}
\affiliation{\Caltech}
\affiliation{\Cambridge}
\affiliation{\UNICAMP}
\affiliation{\CdF}
\affiliation{\FNAL}
\affiliation{\Harvard}
\affiliation{\IIT}
\affiliation{\Indiana}
\affiliation{\IHEP}
\affiliation{\ITEP}
\affiliation{\JMU}
\affiliation{\Lebedev}
\affiliation{\LLL}
\affiliation{\UCL}
\affiliation{\Minnesota}
\affiliation{\Duluth}
\affiliation{\Oxford}
\affiliation{\Pittsburgh}
\affiliation{\RAL}
\affiliation{\USP}
\affiliation{\Carolina}
\affiliation{\Stanford}
\affiliation{\Sussex}
\affiliation{\TexasAM}
\affiliation{\Texas}
\affiliation{\Tufts}
\affiliation{\Washington}
\affiliation{\WandM}
\affiliation{\Wisconsin}

\author{P.~Adamson}
\affiliation{\FNAL}
\affiliation{\UCL}

\author{C.~Andreopoulos}
\affiliation{\RAL}

\author{K.~E.~Arms}
\affiliation{\Minnesota}

\author{R.~Armstrong}
\affiliation{\Indiana}

\author{D.~J.~Auty}
\affiliation{\Sussex}

\author{S.~Avvakumov}
\affiliation{\Stanford}

\author{D.~S.~Ayres}
\affiliation{\ANL}

\author{B.~Baller}
\affiliation{\FNAL}

\author{B.~Barish}
\affiliation{\Caltech}

\author{P.~D.~Barnes~Jr.}
\affiliation{\LLL}

% ...

\collaboration{The MINOS Collaboration}\ \noaffiliation

\date{\today}

\begin{abstract} 

The velocity of a $\sim$3~GeV neutrino beam is measured by comparing detection times at the Near and
Far detectors of the MINOS experiment, separated by 734~km. A total of 473 Far Detector neutrino
events was used to measure $(v-c)/c = 5.1 \pm2.9 \times 10^{-5}$ (at 68\% C.L.).  By correlating the
measured energies of 258 charged-current neutrino events to their arrival times at the Far Detector, a limit is imposed on the neutrino mass of $m_\nu< 50~\textrm{MeV}/c^2$ (99\% C.L.).

\end{abstract}



\maketitle

\section{Introduction}\label{sec:intro}

Investigations of the intrinsic properties of neutrinos and their interactions have led to discoveries such as finite mass, lepton-flavor number violation, and oscillations with large mixing angles \cite{PDBook}.
These surprises warrant careful measurement of other basic neutrino properties such as the relationship between energy and velocity. 

If the mass of the heaviest neutrino is assumed to be 3~eV/$c^2$, the best direct limit on a neutrino mass\cite{Kraus:2004zw,Lobashev:1999tp},
then the relativistic velocity $v$ of a 10~GeV neutrino should satisfy $|v-c|/c\lesssim 10^{-19}$.  Cosmological measurements \cite{Spergel:2003cb} give a mass limit an order of magnitude smaller, implying an even tighter velocity constraint.  However, theories have been proposed to allow some or all neutrinos to travel along ``shortcuts'' off the brane through large extra dimensions \cite{Mohapatra:2006gs}, and thus have apparent velocities different than the speed of light. Some of these theories \cite{Volkov:2006bi,Ammosov:2000kj,Asanov:2000kw} allow $|v-c|/c\sim10^{-4}$ at neutrino energies of a few GeV. Terrestrial neutrino beams could measure an effect of this magnitude.

Earlier terrestrial measurements \cite{Kalbfleisch:1979rm,Alspector:1976kd,Gallas:1994xp} constrained  $|v-c|/c < 4\times10^{-5}$ by comparing the interaction times of muons and muon neutrinos of $E_\nu>30$~GeV created in a 1~ns beam spill over a distance of $\sim$500~m. This work differs in several respects: First, MINOS employs a lower energy beam ($\sim$3~GeV). Second, we measure the absolute transit time of an ensemble of neutrinos, to $<100$~ns accuracy, by comparing neutrino arrival times at the MINOS Near Detector (ND) and Far Detector (FD), separated by a distance of 734~km. Third, we make the unique measurement of comparing the energies of neutrinos in charged-current (CC) interactions to the interaction times in the FD. 

\begin{table} 
\begin{tabular}{ll}
\hline \hline
        \multicolumn{2}{c}{\bf Baseline:}\\ 
	 Distance\footnote{Distance between front face of the ND and the center of the FD.}
	  ND to FD, $L$ & 734\thinspace298.6 $\pm$0.7 m \cite{survey}\\
	 Nominal time of flight, $\tau$         & 2\thinspace449\thinspace356 $\pm$ 2~ns\\[3pt]
	\multicolumn{2}{c}{\bf MINOS Timing System:}\\ 
	GPS Receivers & TrueTime model XL-AK \\ 
	Antenna fiber delay & 1115 ns ND, 5140 ns FD \\ 
	Single Event Time Resolution & $<$40~ns \\ 
	Random Clock Jitter & 100~ns (typical), each site \\[3pt] 
	\multicolumn{2}{c}{\bf Main Injector Parameters:}\\ 
	Main Injector Cycle Time & 2.2~seconds/spill (typical)\\ 
	Main Injector Batches/Spill & 5 or 6 \\ 
	Spill Duration & 9.7~$\mu$s (6 batches)\\
	Batch Duration & 1582~ns \\ 
	Gap Between Batches & 38~ns \\ 
    \hline \hline
	\end{tabular} 
	\caption{Relevant MINOS and NuMI Parameters \label{t:params}}
    
\end{table}

The MINOS detectors \cite{minos-nim,minos:2006rx,Belias:2004bj} are steel-scintillator tracking calorimeters. Planes of 2.54~cm thick steel separate planes made of scintillator strips, 4.1~cm wide and 1~cm thick. The planes are oriented 3.3$^\circ$ from the normal to the beam direction. Strips are aligned orthogonally on adjacent planes to allow three-dimensional reconstruction of event topology. Multi-anode photo-multiplier tubes (PMTs) read out the strips via wavelength-shifting optical fibers.

The time of each PMT hit is recorded by the detector's clock to a precision of 18.8~ns (ND) and 1.6~ns (FD). Although the implementation of these clocks differ, each is synchronized to a Global Positioning System (GPS) receiver to provide absolute Universal Coordinated Time (UTC). The two identical receivers, situated underground, connect to the surface by identical optical transceivers and optical fibers with lengths given in Table~\ref{t:params}. 

The NuMI beam is created by impinging protons from the Fermilab Main Injector onto a graphite target. The secondary mesons are focused by two horns and allowed to decay in a 675~m long decay pipe. The resulting neutrino beam is 93\% $\nu_\mu$, 6\% $\bar{\nu}_\mu$, and 1\% $\nu_e + \bar{\nu}_e$ at the Near detector. After oscillating \cite{minos:2006rx}, the beam at the Far detector is approximately 60\% $\nu_\mu$. The energy spectrum is peaked at approximately 3~GeV, with a long high-energy tail extending to 120~GeV.

The Main Injector accelerates protons to 120~GeV and sends them to NuMI by single-turn extraction. It operates in one of several modes, allowing either 5 or 6 batches of protons per spill. A pulsed dipole magnet extracts protons from the Main Injector. The extraction magnet signal is time-stamped by the ND GPS receiver and defines time of the spill, $t_0$.

\section{Data Selection}

CC $\nu_\mu$ events in the ND are selected using criteria identical to those of Ref.~\cite{minos:2006rx}: events are required to have total reconstructed energy less than 30~GeV, have a vertex contained within a 1~m radius fiducial volume, and be in time with the spill ($\pm$$\sim$7~$\mu$s). A probability-based particle identification parameter removes neutral-current (NC) shower events. This analysis sampled $1.6\times10^6$ ND events, roughly $1/3$ of the first year's data.

The pre-selection of events in the FD requires event times within $\pm$50~$\mu$s of the expected arrival time (assuming a massless neutrino).  Events are accepted if they satisfy one of three selections: $\nu_\mu$ CC events contained in the fiducial volume, neutrino-induced muons from CC interactions in the rock outside the detector (rock-muons), and shower events.

The contained CC event selection is again similar to that of Ref.~\cite{minos:2006rx}; events are required to have a vertex within the fiducial volume, and to have a well-reconstructed track with direction within 53$^\circ$ of that of the beam. In this work, both $\nu_\mu$ and $\bar{\nu}_\mu$ candidates are selected.  Events with tracks penetrating the top of the detector volume are vetoed as possible cosmic-ray contamination. The energy of the CC events is determined by summing hadronic shower energy and muon energy derived from track length or curvature.

Rock-muon events are selected by considering only muons which enter the front face of the detector, to reduce background. The track is required to be contained within the detector volume or to exit the lower half of the detector, to remove background due to cosmic rays reconstructed with the wrong directionality. The track was also required to have a direction within 26$^\circ$ of the beam direction.

Shower events are mostly from NC interactions, but also include CC events from $\nu_e$, and $\nu_\tau$ or $\nu_\mu$ in which no muons are detected. Events with a cluster of hit strips with a total pulse-height greater than approximately $300$~MeV are accepted as shower events. Shower events are required to be contained inside the detector volume, thereby reducing cosmic ray events and random noise.

A total of 473 neutrino-induced FD events were selected, of which 258 were contained $\nu_\mu$ or $\bar{\nu}_\mu$ CC events. By relaxing the selection cuts, the cosmic-ray induced background is estimated to be $<1$ event.

\begin{table}
	\begin{center}
		\begin{tabular}
			{l l r}
			& Description & Uncertainty (68\% C.L.)\\
			\hline 
			A & Distance between detectors   & 2~ns \\ 
			B & ND Antenna fiber length      & 27~ns\\ 
			C & ND electronics latencies     & 32~ns\\
			D & FD Antenna fiber length      & 46~ns\\
			E & FD electronics latencies     & 3~ns\\
			F & GPS and transceivers         & 12~ns\\
			G & Detector readout differences &  9~ns\\
			\hline 
			  & Total (Sum in quadrature) & 64~ns \\ 
			\hline 
		\end{tabular}
	\caption{Sources of uncertainty in $\nu$ relative time measurement. \label{t:sys_simple}} 
	\end{center}
\end{table}

\section{Neutrino Event Timing}

The time of a neutrino interaction in the ND is taken as time of the earliest scintillator hit, $t_{ND}$. This time is compared to the time of extraction magnet signal, $t_0$, and corrected for known timing delays: $t_1 = t_{ND} - t_0 - d_{ND}$. Similarly, for FD events, $t_2 = t_{FD} - t_0 - d_{FD}$.


The corrections $d_{ND}$ and $d_{FD}$ incorporate known offsets and delays due to readout time, electronic latency, and GPS antenna fiber delays. Test-stand measurements were used to find the magnitude of each offset. Table~\ref{t:sys_simple} summarizes the uncertainties on these corrections. The uncertainty on the net correction $|d_{ND}-d_{FD}|$ was determined to be $\pm$64~ns at a 68\% C.L. The delay of optical fibers that run between the surface antennas and the underground GPS receivers created the largest uncertainties; these uncertainties were estimated from the dispersion of multiple independent measurements of the fiber delays. For example, the delay of a Far Detector fibre was independently measured with four instruments: an Optical Wavelength Labs BOLT-NL, an Aligent E6000B Optical Time Domain Reflectometer (OTDR), an EXFO FTB-300 OTDR, and a custom-built test apparatus. The results of these four measurements all differed, with an RMS of 46~ns, roughly 1\% of the delay.

\begin{figure}[tb] \begin{center}
\includegraphics[width=0.95\columnwidth]{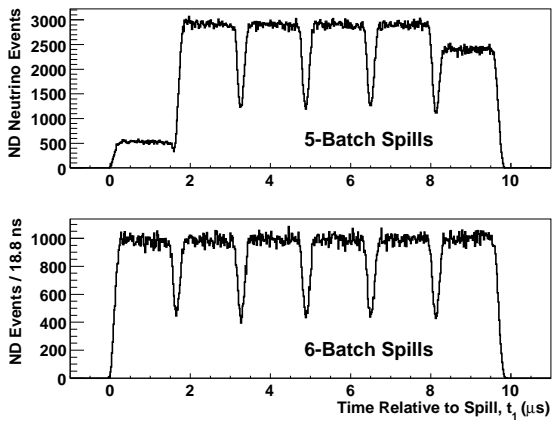}
\caption{Neutrino event time distribution measured at the MINOS Near
  Detector. The top plot corresponds to events in 5-batch spills $P_1^5(t_1)$ while the
  bottom plot corresponds to 6-batch spills $P_1^6(t_1)$.}
\label{f:nd}
\end{center}\end{figure}

If the pulse of neutrinos were instantaneous, the deviation from the expected time-of-flight $\tau$ could be measured as $\delta = (t_2-t_1) - \tau$. However, the NuMI beam pulse is 9.7~$\mu$s long, with an intensity time-profile consisting of 6 batches with short gaps in between.  The relative intensities of these batches, and the shape of the gaps, are due to the proton intensity profile of the NuMI beam. Two running modes, `5-batch' and `6-batch' are distinguished on a pulse-by-pulse basis. (The `5-batch' mode includes two types of spill, where either the first or last batch are not delivered to NuMI.) The ND measures this intensity profile with neutrino interactions. This measurement calibrates the neutrino time relative to the extraction fire signal.  This profile is shown as Figure~\ref{f:nd} and is represented as probability density 
functions (PDFs) $P_1^5(t_1)$ and $P_1^6(t_1)$.

The arrival time distribution of neutrinos at the FD is similar, but the relative jitter of the two GPS clocks further degrades the time resolution. These clocks have a maximum error of $\pm$200~ns relative to UTC, with a typical error of 100~ns.  The uncorrelated jitter of two clocks, in addition to detector time resolution, gives a total relative (FD/ND) time uncertainty of $\sigma=150$~ns. We therefore compose PDFs of the expected FD neutrino arrival time distribution $P^5_2$ and $P^6_2$: 

\begin{eqnarray}
\nonumber P^n_2\left(t_2\right) = \int 
  \frac{1}{\sigma\sqrt{2\pi}} \exp\left(-\frac{(t_2-t')^2}{2\sigma^2}\right)\\
 \times P^n_1 \left( t' \right) dt' \qquad (n=5,6)\label{eq:p2}
\end{eqnarray}

\begin{figure}[t] \begin{center}
\includegraphics[width=0.95\columnwidth]{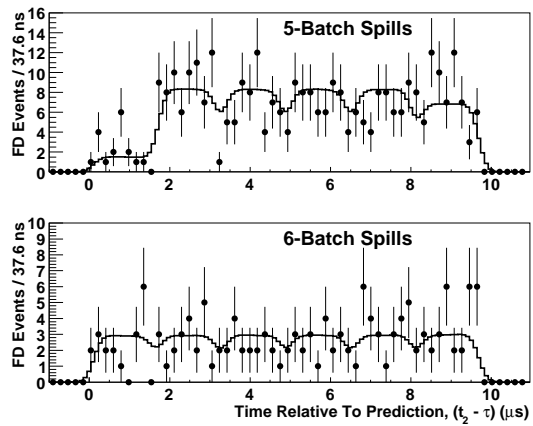}
\caption{\label{f:tof_data} Time distribution of FD events relative to
prediction after fitting the time-of-flight. The top plot shows events in 5-batch spills, the bottom 6-batch spills. The normalized expectation curves $P^5_2(t)$ and $P^6_2(t)$ are shown as the solid lines.}
\end{center} \end{figure}

The resulting PDF describes the predicted time distribution at the FD. The time of each event in the FD ($t_2^i$) was compared to this PDF.  The deviation $\delta$ from the expected time was found by maximizing a log-likelihood function ($L$), summing each event ($i$) in the 5- and 6-batch data:

\begin{equation}
L = \sum_i \ln \ P_2 \left( t_2^i - \tau -\delta \right) .\label{eq:L}
\end{equation}
The distribution of measured FD times is shown in Figure~\ref{f:tof_data}, along with the predicted distribution for the best fit value of $\delta$. The deviation was found to be $\delta = -126 \pm 32 \textrm{(stat.)} \pm 64 \textrm{(sys.)}$~ns at a 68\% confidence limit (C.L.). The systematic uncertainty is due to timing offsets shown in Table~\ref{t:sys_simple}. The goodness-of-fit probability was determined by Monte Carlo (MC) to be 10\%, and the likelihood is Gaussian in $\delta$.

\section{Relativistic Mass Measurement} \label{sec:mass_ana}

If the neutrino had a relativistic mass $m_\nu$ and total energy $E_\nu$, the time of flight would be:
\begin{equation}
T_{m_\nu}(E_\nu) = \frac{\tau}{\sqrt{1-\left(\frac{m_\nu c^2}{E_\nu}\right)^2}},
\end{equation}

where $\tau$ is the time of flight of a massless particle. For contained $\nu_\mu$ and $\bar{\nu}_\mu$ CC events, the MINOS detectors measure the neutrino energy, allowing the hypothesis of a non-zero mass to be tested by measuring the arrival time as a function of $E_\nu$.

The measured event times are fitted to a model with $m_\nu$ as a free parameter. For symmetry, the fit allowed $m_\nu$ to become negative; positive masses indicated retarded arrival times, while negative masses were interpreted as advanced arrival times, i.e. $T_{-m_\nu} \rightarrow \tau+(\tau-T_{m_\nu})$. The detector energy response $R(E_\nu,E_{\rm reco})$ was parameterized by a PDF derived from MC, where $E_{\rm reco}$ is the energy reconstructed in the detector. The true energies $E^i_\nu$ are unknown, and so are fitted as a set of 258 parameters constrained by $R$ and $E_{\rm reco}$. At the peak beam energy, $R$ is approximately Gaussian in $E_\nu-E_{\rm reco}$ with a width of $\sim$30\%.

The offset $\delta$ is taken as a parameter constrained by the earlier systematic measurements as a Gaussian about zero with $\sigma_\delta = 64$~ns. A log-likelihood is constructed using the expected arrival time PDF $P_2(t_2)$, the arrival times $t^i_2$, the fitted true energies $E^i_\nu$. 
\begin{equation}
L =  \frac{\delta^2}{2 \sigma_\delta^2}
+ \sum_i \left[ \ln  P_2\left(t^i_2 - T_{m_\nu}(E^i_\nu) -\delta \right) 
+ \ln R\left(E^i_\nu,E^i_{\rm reco}\right) \right]
\label{eq:massL}
\end{equation}
The result of the fit was $m_\nu=17^{+13}_{-28}$(stat.)~MeV$/c^2$ at a 68\% C.L. The likelihood function is non-Gaussian; the 99\% C.L is $m_\nu=17^{+33}_{-46}$(stat.)~MeV$/c^2$. The best fit gave $\delta=-46$~ns and a goodness-of-fit probability of 8\%. 

\begin{figure}[tb] \begin{center}
\includegraphics[width=0.95\columnwidth]{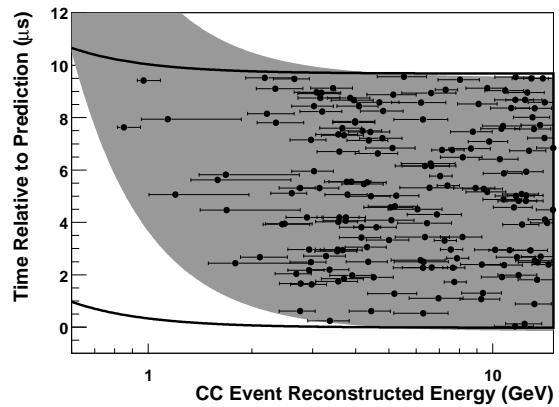}
\caption{The time and reconstructed energy for contained $\nu_\mu$ charged current events. The points show the measured times of events and reconstructed energy $E_{\rm reco}$. The horizontal error bars indicate the $\sim$1$\sigma$ energy uncertainty. The gray filled region indicates the allowed range of times predicted by a neutrino with $m_\nu=50$~MeV$/c^2$. The solid lines indicate the allowed range predicted $m_\nu=17$~MeV$/c^2$.}
\label{f:mass_data}
\end{center} 
\end{figure}

The uncertainty on the energy resolution of the FD dominates the systematic uncertainty. The response $R\left(E^i_\nu,E^i_{\rm reco}\right)$ was found by comparing the reconstructed neutrino energy with the input neutrino energy in a MC simulation, with events weighted for an oscillated neutrino beam, with $\Delta m_{23}^2=0.0027\ \textrm{eV}^2, \sin^22\theta_{23}=1.0$~\cite{minos:2006rx}. To estimate the systematic uncertainty, the detector response $R$ was varied by (a) changing the expected neutrino energy distribution by varying oscillations parameters within the allowed range of Ref.~\cite{minos:2006rx} (b) increasing the NC contamination of the CC sample by $\pm50$\%, (c) changing the shower energy scale by $\pm11\%$, and (d) changing the muon energy scale by $\pm$2\%. For each change, $R$ was evaluated and the data re-analyzed. We incorporate these systematics by simply taking the extremum limits on $m_\nu$ from all of these trials, obtaining a final result of $m_\nu=17^{+33}_{-56} \textrm{(stat.+sys.)}$~MeV$/c^2$, 99\% C.L. The limiting case of $m_\nu=50$~MeV$/c^2$ is shown graphically in Figure~\ref{f:mass_data}, which shows the data for events with energies less than 10~GeV. Neutrinos consistent with this mass fall inside the shaded region. 

In practice, this method uses the high-energy events to constrain $\delta$, and uses the lowest-energy events to constrain the relativistic neutrino mass.  If the constraint on $\delta$ is removed, a free fit gives $m_\nu=14_{-48}^{+42} \textrm{(stat.)}$~MeV$/c^2$ and $\delta=-99\pm140\ (stat.)$~ns at a 99\% C.L., with a probability of fit of 10\%.

\section{Conclusions}
By measuring the arrival time of 473 contained CC, rock-muon, and NC events as measured by the MINOS GPS clocks, the deviation from the expected time at the Far Detector was found to be
$$ \delta = -126 \pm 32 \ \textrm{(stat.)} \pm 64 \ \textrm{(sys.)} \ \textrm{ns} \quad \textrm{68\%\ C.L.} $$
By comparing to the nominal time of flight $\tau$, we interpret this as a neutrino velocity of $v=L/(\tau+\delta)$ to satisfy
$$ \frac{(v-c)}{c} = \frac{-\delta}{\tau+\delta} = 5.1 \pm 2.9 \textrm{(stat.+sys.)} \times 10^{-5} \  \textrm{68\%\ C.L.} $$
for neutrinos of $\sim$3~GeV. This measurement is consistent with the speed of light to less than 1.8$\sigma$.  The corresponding 99\% confidence limit on the speed of the neutrino is
$ -2.4 \times 10^{-5} < (v-c)/c < 12.6 \times 10^{-5} $.

This measurement has implicitly assumed that the $m_2$ and $m_3$ neutrino mass eigenstates that comprise the beam are traveling at the same velocity. This assumption is borne out in observing that the arrival times at the Far detector match the expectation distribution.  Indeed, if the two eigenstates were to travel at velocities differing by as little as $\Delta v/v \gtrsim 4\times10^{-7}$, the short $\sim$1~ns bunches would separate in transit and thus decohere, changing or destroying oscillation effects at the Far detector.  

Besides the novelty of the technique, this measurement is unique in that it probes the 1-30~GeV region of neutrino energy not measured by previous experiments. The measurements described in Refs. \cite{Kalbfleisch:1979rm,Alspector:1976kd} reached a sensitivity slightly better than this work, but only for neutrinos of $\sim25$~GeV and higher. The most sensitive test of neutrino velocity was achieved by comparing\footnote{The SN1987A measurement is predicated on the theoretical assumption that neutrinos and photons are emitted within three hours of each other.} the arrival times of neutrinos \cite{Hirata:1987hu,Bionta:1987qt} and photons from SN1987a, which achieved a sensitivity of $|v-c|/c<2\times10^{-9}$ \cite{Stodolsky:1987vd,Longo:1987gc}, four orders of magnitude better than the terrestrial measurements, but only for neutrinos of energy $\sim10$~MeV.  In principle, neutrino velocity could be a strong function of energy. Our measurement constrains this previously untested regime.

By using the arrival time and reconstructed energies of 258 contained CC $\nu_\mu$ and $\bar{\nu}_\mu$ events, a limit was found on the relativistic mass of the neutrino:
$$ m_\nu < 50 \ \textrm{MeV}/c^2 \ \textrm{(stat.+sys.) \quad 99\%\ C.L.} $$
The method of relating time-of-flight to neutrino energy is a new technique made available by the MINOS calorimeters.  With the entire MINOS data sample we anticipate a factor of 10 increase in statistics.  This will improve the sensitivity of this measurement to as low as $10$~MeV$/c^2$.  

Direct mass measurement yields much tighter constraint of neutrino mass \cite{Kraus:2004zw,Lobashev:1999tp}, but the method of measuring mass by time-of-flight can additionally be viewed as a test of the relativistic energy-velocity relation. If the relation holds, the mass measured by time-of-flight should be consistent with direct measurements. Alternate theories, such as those suggested by Refs.~\cite{Volkov:2006bi,Ammosov:2000kj,Asanov:2000kw} could be tested by these data. 

\section{Acknowledgments}
We thank the Fermilab staff and the technical staffs of the participating institutions for their vital contributions. This work was supported by the U.S. Department of Energy, the U.S. National Science Foundation, the U.K. Particle Physics and Astronomy Research Council, the State and University of Minnesota, the Office of Special Accounts for Research Grants of the University of Athens, Greece, FAPESP (Rundacao de Amparo a Pesquisa do Estado de Sao Paulo), CNPq (Conselho Nacional de Desenvolvimento Cientifico e Tecnologico) in Brazil, and the computational resources of the AVIDD cluster at Indiana University. We gratefully acknowledge the Minnesota Department of Natural Resources for their assistance and for allowing us access to the facilities of the Soudan Underground Mine State Park. We also thank the crew of the Soudan Underground Physics laboratory for their tireless work in building and operating the MINOS detector.


 \bibliography{minostof}

\begin{thebibliography}{20}
\expandafter\ifx\csname natexlab\endcsname\relax\def\natexlab#1{#1}\fi
\expandafter\ifx\csname bibnamefont\endcsname\relax
  \def\bibnamefont#1{#1}\fi
\expandafter\ifx\csname bibfnamefont\endcsname\relax
  \def\bibfnamefont#1{#1}\fi
\expandafter\ifx\csname citenamefont\endcsname\relax
  \def\citenamefont#1{#1}\fi
\expandafter\ifx\csname url\endcsname\relax
  \def\url#1{\texttt{#1}}\fi
\expandafter\ifx\csname urlprefix\endcsname\relax\def\urlprefix{URL }\fi
\providecommand{\bibinfo}[2]{#2}
\providecommand{\eprint}[2][]{\url{#2}}

\bibitem[{\citenamefont{{Yao} et~al.}(2006)}]{PDBook}
\bibinfo{author}{\bibfnamefont{W.-M.} \bibnamefont{{Yao}}}
  \bibnamefont{et~al.}, \bibinfo{journal}{{Journal of Physics G}}
  \textbf{\bibinfo{volume}{33}} (\bibinfo{year}{2006}).

\bibitem[{\citenamefont{Kraus et~al.}(2005)}]{Kraus:2004zw}
\bibinfo{author}{\bibfnamefont{C.}~\bibnamefont{Kraus}} \bibnamefont{et~al.},
  \bibinfo{journal}{Eur. Phys. J.} \textbf{\bibinfo{volume}{C40}},
  \bibinfo{pages}{447} (\bibinfo{year}{2005}), \eprint{hep-ex/0412056}.

\bibitem[{\citenamefont{Lobashev et~al.}(1999)}]{Lobashev:1999tp}
\bibinfo{author}{\bibfnamefont{V.~M.} \bibnamefont{Lobashev}}
  \bibnamefont{et~al.}, \bibinfo{journal}{Phys. Lett.}
  \textbf{\bibinfo{volume}{B460}}, \bibinfo{pages}{227} (\bibinfo{year}{1999}).

\bibitem[{\citenamefont{Spergel et~al.}(2003)}]{Spergel:2003cb}
\bibinfo{author}{\bibfnamefont{D.~N.} \bibnamefont{Spergel}}
  \bibnamefont{et~al.} (\bibinfo{collaboration}{WMAP}),
  \bibinfo{journal}{Astrophys. J. Suppl.} \textbf{\bibinfo{volume}{148}},
  \bibinfo{pages}{175} (\bibinfo{year}{2003}).

\bibitem[{\citenamefont{Mohapatra and Smirnov}(2006)}]{Mohapatra:2006gs}
\bibinfo{author}{\bibfnamefont{R.~N.} \bibnamefont{Mohapatra}}
  \bibnamefont{and} \bibinfo{author}{\bibfnamefont{A.~Y.}
  \bibnamefont{Smirnov}}, \bibinfo{journal}{Ann. Rev. Nucl. Part. Sci.}
  \textbf{\bibinfo{volume}{56}}, \bibinfo{pages}{569} (\bibinfo{year}{2006}),
  \eprint{hep-ph/0603118}.

\bibitem[{\citenamefont{Volkov}(2006)}]{Volkov:2006bi}
\bibinfo{author}{\bibfnamefont{G.~G.} \bibnamefont{Volkov}},
  \bibinfo{howpublished}{Presented at 2nd Symposium on Neutrinos and Dark
  Matter in Nuclear Physics (NDM06)} (\bibinfo{year}{2006}).

\bibitem[{\citenamefont{Ammosov and Volkov}(2000)}]{Ammosov:2000kj}
\bibinfo{author}{\bibfnamefont{V.}~\bibnamefont{Ammosov}} \bibnamefont{and}
  \bibinfo{author}{\bibfnamefont{G.}~\bibnamefont{Volkov}}
  (\bibinfo{year}{2000}), \eprint{DFPD-00-TH-39, hep-ph/0008032}.

\bibitem[{\citenamefont{Asanov}(2000)}]{Asanov:2000kw}
\bibinfo{author}{\bibfnamefont{G.~S.} \bibnamefont{Asanov}}
  (\bibinfo{year}{2000}), \eprint{hep-ph/0009305}.

\bibitem[{\citenamefont{Kalbfleisch et~al.}(1979)\citenamefont{Kalbfleisch,
  Baggett, Fowler, and Alspector}}]{Kalbfleisch:1979rm}
\bibinfo{author}{\bibfnamefont{G.~R.} \bibnamefont{Kalbfleisch}},
  \bibinfo{author}{\bibfnamefont{N.}~\bibnamefont{Baggett}},
  \bibinfo{author}{\bibfnamefont{E.~C.} \bibnamefont{Fowler}},
  \bibnamefont{and}
  \bibinfo{author}{\bibfnamefont{J.}~\bibnamefont{Alspector}},
  \bibinfo{journal}{Phys. Rev. Lett.} \textbf{\bibinfo{volume}{43}},
  \bibinfo{pages}{1361} (\bibinfo{year}{1979}).

\bibitem[{\citenamefont{Alspector et~al.}(1976)}]{Alspector:1976kd}
\bibinfo{author}{\bibfnamefont{J.}~\bibnamefont{Alspector}}
  \bibnamefont{et~al.}, \bibinfo{journal}{Phys. Rev. Lett.}
  \textbf{\bibinfo{volume}{36}}, \bibinfo{pages}{837} (\bibinfo{year}{1976}).

\bibitem[{\citenamefont{Gallas et~al.}(1995)}]{Gallas:1994xp}
\bibinfo{author}{\bibfnamefont{E.}~\bibnamefont{Gallas}} \bibnamefont{et~al.}
  (\bibinfo{collaboration}{FMMF}), \bibinfo{journal}{Phys. Rev.}
  \textbf{\bibinfo{volume}{D52}}, \bibinfo{pages}{6} (\bibinfo{year}{1995}).

\bibitem[{\citenamefont{Bocean}(1999)}]{survey}
\bibinfo{author}{\bibfnamefont{V.}~\bibnamefont{Bocean}}, in
  \emph{\bibinfo{booktitle}{Pro. 6th Int. Workshop on Acc. Alignment}}
  (\bibinfo{organization}{IWAA99}, \bibinfo{address}{Grenoble, France},
  \bibinfo{year}{1999}).

\bibitem[{\citenamefont{Michael et~al.}(2007)}]{minos-nim}
\bibinfo{author}{\bibfnamefont{D.~G.} \bibnamefont{Michael}}
  \bibnamefont{et~al.} (\bibinfo{collaboration}{MINOS}), \bibinfo{journal}{To
  be submitted to Nucl. Inst. Meth.}  (\bibinfo{year}{2007}).

\bibitem[{\citenamefont{Michael et~al.}(2006)}]{minos:2006rx}
\bibinfo{author}{\bibfnamefont{D.~G.} \bibnamefont{Michael}}
  \bibnamefont{et~al.} (\bibinfo{collaboration}{MINOS}),
  \bibinfo{journal}{Phys. Rev. Lett.} \textbf{\bibinfo{volume}{97}},
  \bibinfo{pages}{191801} (\bibinfo{year}{2006}).

\bibitem[{\citenamefont{Belias et~al.}(2004)}]{Belias:2004bj}
\bibinfo{author}{\bibfnamefont{A.}~\bibnamefont{Belias}} \bibnamefont{et~al.},
  \bibinfo{journal}{IEEE Trans. Nucl. Sci.} \textbf{\bibinfo{volume}{51}},
  \bibinfo{pages}{451} (\bibinfo{year}{2004}).

\bibitem[{\citenamefont{Pas et~al.}(2005)\citenamefont{Pas, Pakvasa, and
  Weiler}}]{Pas:2005rb}
\bibinfo{author}{\bibfnamefont{H.}~\bibnamefont{Pas}},
  \bibinfo{author}{\bibfnamefont{S.}~\bibnamefont{Pakvasa}}, \bibnamefont{and}
  \bibinfo{author}{\bibfnamefont{T.~J.} \bibnamefont{Weiler}},
  \bibinfo{journal}{Phys. Rev.} \textbf{\bibinfo{volume}{D72}},
  \bibinfo{pages}{095017} (\bibinfo{year}{2005}).

\bibitem[{\citenamefont{Hirata et~al.}(1987)}]{Hirata:1987hu}
\bibinfo{author}{\bibfnamefont{K.}~\bibnamefont{Hirata}} \bibnamefont{et~al.}
  (\bibinfo{collaboration}{KAMIOKANDE-II}), \bibinfo{journal}{Phys. Rev. Lett.}
  \textbf{\bibinfo{volume}{58}}, \bibinfo{pages}{1490} (\bibinfo{year}{1987}).

\bibitem[{\citenamefont{Bionta et~al.}(1987)}]{Bionta:1987qt}
\bibinfo{author}{\bibfnamefont{R.~M.} \bibnamefont{Bionta}}
  \bibnamefont{et~al.}, \bibinfo{journal}{Phys. Rev. Lett.}
  \textbf{\bibinfo{volume}{58}}, \bibinfo{pages}{1494} (\bibinfo{year}{1987}).

\bibitem[{\citenamefont{Stodolsky}(1988)}]{Stodolsky:1987vd}
\bibinfo{author}{\bibfnamefont{L.}~\bibnamefont{Stodolsky}},
  \bibinfo{journal}{Phys. Lett.} \textbf{\bibinfo{volume}{B201}},
  \bibinfo{pages}{353} (\bibinfo{year}{1988}).

\bibitem[{\citenamefont{Longo}(1988)}]{Longo:1987gc}
\bibinfo{author}{\bibfnamefont{M.~J.} \bibnamefont{Longo}},
  \bibinfo{journal}{Phys. Rev. Lett.} \textbf{\bibinfo{volume}{60}},
  \bibinfo{pages}{173} (\bibinfo{year}{1988}).

\end{thebibliography}

 \end{document}